# The Structure of Collaborative Tagging Systems

Scott A. Golder and Bernardo A. Huberman
Information Dynamics Lab, HP Labs
{ scott.golder , bernardo.huberman }@hp.com

**ABSTRACT**

Collaborative tagging describes the process by which many users add metadata in the form of keywords to shared content. Recently, collaborative tagging has grown in popularity on the web, on sites that allow users to tag bookmarks, photographs and other content. In this paper we analyze the structure of collaborative tagging systems as well as their dynamical aspects. Specifically, we discovered regularities in user activity, tag frequencies, kinds of tags used, bursts of popularity in bookmarking and a remarkable stability in the relative proportions of tags within a given url. We also present a dynamical model of collaborative tagging that predicts these stable patterns and relates them to imitation and shared knowledge.

**KEYWORD LIST**

Collaborative tagging, folksonomy, Del.icio.us, bookmarks, web, sharing.

## 1. INTRODUCTION

Marking content with descriptive terms, also called keywords or tags, is a common way of organizing content for future navigation, filtering or search. Though organizing electronic content this way is not new, a collaborative form of this process, which has been given the name "tagging" by its proponents, is gaining popularity on the web.

Document repositories or digital libraries often allow documents in their collections to be organized by assigned keywords. However, traditionally such categorizing or indexing is either performed by an authority, such as a librarian, or else derived from the material provided by the authors of the documents (Rowley 1995). In contrast, collaborative tagging is the practice of allowing anyone – especially consumers – to freely attach keywords or tags to content. Collaborative tagging is most useful when there is nobody in the "librarian" role or there is simply too much content for a single authority to classify; both of these traits are true of the web, where collaborative tagging has grown popular.

This kind of collaborative tagging offers an interesting alternative to current efforts at semantic web ontologies (Shirky 2005) which have been a focus of research by a number of groups (e.g. Doan, Madhavan, Domingos & Halevy 2002).

A number of now-prominent web sites feature collaborative tagging. Typically, such sites allow users to publicly tag and share content, so that they can not only categorize information for themselves, they can browse the information categorized by others. There is therefore at once both personal and public aspects to collaborative tagging systems. In some sites, collaborative tagging is also known as "folksonomy," short for "folk taxonomy;" however, there is some debate whether this term is accurate (Mathes 2004), and so we avoid using it here.

Del.icio.us, the site on which we performed our analysis, allows for the collaborative tagging of shared website bookmarks. Yahoo's MyWeb does this as well, and CiteULike and Connotea do the same for references to academic publications. Some services allow users to tag, but only content they own, for example, Flickr for photographs and Technorati for weblog posts. Though these two sites do not, strictly speaking, support *collaborative* tagging, we mention them to illustrate the growth of tagging in a variety of media.

In this paper we analyze the structure of collaborative tagging systems as well as their dynamical aspects. Specifically, through the study of the collaborative tagging system Delicious, we are able to discover regularities in user activity, tag frequencies, kinds of tags used and bursts of popularity in bookmarking, as well as a remarkable stability in the relative proportions of tags within a given url. We also present a dynamical model of collaborative tagging that predicts these stable patterns and relates them to imitation and shared knowledge. We conclude with a discussion of potential uses of the data that users of these systems collaboratively generate.

## 2. TAGGING AND TAXONOMY

Proponents of collaborative tagging, typically in the weblogging community, often contrast tagging-based systems from taxonomies. While the latter are hierarchical and exclusive, the former are non-hierarchical and inclusive. Familiar taxonomies include the Linnaean system of classifying living things, the Dewey Decimal classification for libraries, and computer file systems for organizing electronic files. In such systems, each animal, book, file and so on, is in one unambiguous category which is in turn within a yet more general one. For example, lions and tigers fall in the genus *panthera*, and domestic cats in the genus *felis*, but *panthera* and *felis* are both part of family *felidae*, of which lions, tigers and domestic cats are all part. Similarly, books on Africa's geography are in the Dewey Decimal system category 916 and books on South America's in 918, but both are subsumed by the 900 category, covering all topics in geography.

In contrast, tagging is neither exclusive nor hierarchical and therefore can in some circumstances have an advantage over hierarchical taxonomies. For example, consider a hypothetical researcher who downloads an article about cat species native to Africa. If the researcher wanted to organize all her downloaded articles in a hierarchy of folders, there are several hypothetical options, of which we consider four:

1. `c:\articles\cats`           all articles on cats
2. `c:\articles\africa`         all articles on Africa
3. `c:\articles\africa\cats`    all articles on African cats
4. `c:\articles\cats\africa`    all articles on cats from Africa

Each choice reflects a decision about the relative importance of each characteristic. Folder names and levels are in themselves informative, in that, like tags, they describe the information held within them (Jones et al. 2005). Folders like 1. and 2. make central the fact that the folders are about "`cats`" and "`africa`" respectively, but elide all information about the other category. 3. and 4. organize the files by both categories, but establish the first as primary or more salient, and the second as secondary or more specific. However, looking in 3. for a file in 4. will be fruitless, and so checking multiple locations becomes necessary.

Despite these limitations, there are several good reasons to impose such a hierarchy. Though there can be too many folders in a hierarchy, especially one created haphazardly, an efficiently organized file hierarchy neatly and unambiguously bounds a folder's contents. Unlike a keyword-based search, wherein the seeker cannot be sure that a query has returned all relevant items, a folder hierarchy assures the seeker that all the files it contains are in one stable place.

In contrast to a hierarchical file system, a non-exclusive, flat tagging system could, unlike the system described above, identify such an article as being about a great variety of things simultaneously, including `africa` and `cats`, as well as `animals` more generally, and `cheetahs`, more specifically.

Like a Venn diagram, the set of all the items marked `cats` and those marked `africa` would intersect in precisely one way, namely, those documents that are tagged as being about African cats. Even this is not perfect, however. For example, a document tagged only `cheetah` would not be found in the intersection of `africa` and `cats`, though it arguably ought to; like the foldering example above, a seeker may still need to search multiple locations.

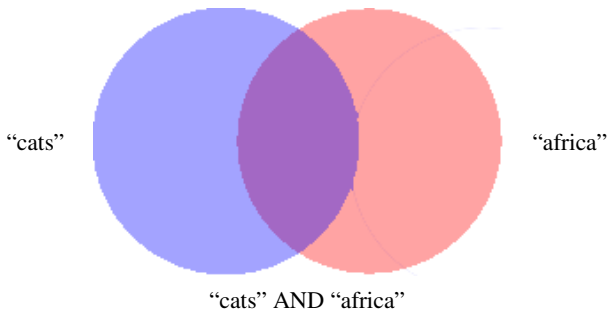

**Figure 1. A Venn diagram showing the intersection of "cats" and "africa".**

Looking at it another way, tagging is like filtering; out of all the possible documents (or other items) that are tagged, a filter (i.e. a tag) returns only those items tagged with that tag. Depending on the implementation and query, a tagging system can, instead of providing the intersection of tags (thus, filtering), provide the union of tags; that is, all the items tagged with *any* of the given tags, rather than *all* of them. From a user perspective, navigating a tag system is similar to conducting keyword-based searches; regardless of the implementation, users are providing salient, descriptive terms in order to retrieve a set of applicable items.

## 2.1 Semantic and Cognitive Aspects of Classification

Both tagging systems and taxonomies are beset by many problems that exist as a result of the necessarily imperfect, yet natural and evolving process of creating semantic relations between words and their referents. Three of these problems are polysemy, synonymy, and basic level variation.

A polysemous word is one that has many ("poly") related senses ("semy"). For example, a "window" may refer to a hole in the wall, or to the pane of glass that resides within it (Pustejovsky 1995). In practice, polysemy dilutes query results by returning related but potentially inapplicable items. Superficially, polysemy is similar to homonymy, where a word has multiple, unrelated meanings. However, homonymy is less a problem because homonyms can be largely ruled out in a tag-based search through the addition of a related term with which the unwanted homonym would not appear. There are, of course, cases where homonyms are semantically related but not polysemous; for example, searching for employment at Apple may be problematic because of conflicts with the CEO's surname.

Synonymy, or multiple words having the same or closely related meanings, presents a greater problem for tagging systems because inconsistency among the terms used in tagging can make it very difficult for one to be sure that all the relevant items have been found. It is difficult for a tagger to be consistent in the terms chosen for tags; for example, items about television may be tagged either `television` or `tv`. This problem is compounded in a collaborative system, where all taggers either need to widely agree on a convention, or else accept that they must issue multiple or more complex queries to cover many possibilities. Synonymy is a significant problem because it is impossible to know how many items "out there" one would have liked one's query to have retrieved, but didn't.

Relatedly, plurals and parts of speech and spelling can stymie a tagging system. For example, if tags `cat` and `cats` are distinct, then a query for one will not retrieve both, unless the system has the capability to perform such replacements built into it.

Reflecting the cognitive aspect of hierarchy and categorization, the "basic level" problem is that related terms that describe an item vary along a continuum of specificity ranging from very general to very specific; as discussed above, `cat`, `cheetah` and `animal` are all reasonable ways to describe a particular entity. The problem lies in the fact that different people may consider terms at different levels of specificity to be most useful or appropriate for describing the item in question. The "basic level," as opposed to superordinate (more general) and subordinate (more specific) levels, is that which is most directly related to humans' interactions with them (Tanaka & Taylor 1991). For most people, the basic level for felines would be "cat," rather than "animal" or "siamese" or "persian." Experiments demonstrate that, when asked to identify dogs and birds, subjects used "dog" and "bird" more than "beagle" or "robin," and when asked whether an item in a picture is an X, subjects responded more quickly when X was a "basic" level (Tanaka & Taylor 1991). These experiments demonstrate general agreement across subjects.

There is, however, systematic variation across individuals in what constitutes a basic level. Expertise plays a role in defining what level of specificity an individual treats as "basic." For example, in the bird and dog experiments, subjects expert in one of the two domains demonstrated basic levels that were at levels of greater specificity than those without domain expertise; a dog expert might consider "beagle" a basic level where a bird expert might have "dog" and a bird expert "robin" where a dog expert has "bird" (Tanaka & Taylor 1991).

The underlying factor behind this variation may be that basic levels vary in specificity to the degree that such specificity makes a difference in the lives of the individual. A dog expert has not only the skill but also the need to differentiate beagles from poodles, for example. Like variation in expertise, variations in other social or cultural categories likely yield variations in basic levels.

For the purposes of tagging systems, however, conflicting basic levels can prove disastrous, as documents tagged `perl` and `javascript` may be too specific for some users, while a document tagged `programming` may be too general for others.

Tagging is fundamentally about sensemaking. Sensemaking is a process in which information is categorized and labeled and, critically, through which meaning emerges (Weick, Sutcliffe & Obstfeld forthcoming). Recall that "basic levels" are related to the way in which humans interact with the items at those levels (Tanaka & Taylor 1991); when one interacts with the outside world, one makes sense of the things one encounters by categorizing them and ascribing meaning to them. However, in practice, categories are often not well defined and their boundaries exhibit vagueness (Labov 1973). Items often lie between categories or equally well in multiple categories. The lines one ultimately draws for oneself reflect one's own experiences, daily practices, needs and concerns.

Sensemaking is also influenced by social factors (Weick et al. forthcoming). Because many experiences are shared with others and may be nearly universal within a culture or community, similar ways of organizing and sensemaking do result; after all, societies are able to collectively organize knowledge and coordinate action. Additionally, collective sensemaking is subject to conflict between the participating actors, where different opinions and perspectives can clash and power struggles to determine the terms of the debate can ensue (Weick et al. forthcoming).

Collective tagging, then, has the potential to exacerbate the problems associated with the fuzziness of linguistic and cognitive boundaries. As all taggers' contributions collectively produce a larger classification system, that system consists of idiosyncratically personal categories as well as those that are widely agreed upon. However, there is also opportunity to learn from one another through sharing and organizing information.

# 3. DELICIOUS DYNAMICS

Del.icio.us, or Delicious, is a collaborative tagging system for web bookmarks that its creator, Joshua Schachter, calls a "social bookmarks manager" (Delicious n.d.).

We analyzed data from Delicious to uncover patterns among users, tags and URLs. We briefly describe Delicious and analyze tags in this section, and analyze bookmarks and URLs in the following section. Finally, we discuss the value of this collaboratively generated data.

Much in the same way users save bookmarks within their browsers, they can save bookmarks in Delicious, instead; the benefit of doing so is that once one's bookmarks are on the web, they are accessible from any computer, not just the user's own browser. This is helpful if one uses multiple computers, at home, work, school, and so on, and is touted as one of Delicious' main features.

Once users have created accounts, they may then begin bookmarking web pages; each bookmark records the web page's URL and its title, as well as the time at which the bookmark is created. Users can also choose to tag the bookmark with multiple tags, or keywords, of their choice. Each user has a personal page on which their bookmarks are displayed; this page is located at *http://del.icio.us/username*. On this page, all the bookmarks the user has ever created are displayed in reverse-chronological order along with a list of all the tags the user has ever given to a bookmark. By selecting a tag, one can filter one's bookmark list so that only bookmarks with that tag are displayed.

Delicious is considered "social" because, not only can one see one's own bookmarks, one can also see all of every other user's bookmarks. The front page of Delicious shows several of the most recently added bookmarks, including the tags given to them, who created them, and how many other people have that bookmark in common. There is also a "popular" page, which shows the same information for the URLs that are currently the most popular. One can also see any other user's personal page and filter it by tag, much in the way one can one's own.

Through others' personal pages and the "popular" page, users can get a sense of what other people find interesting. By browsing specific people and tags, users can find websites that are of interest to them and can find people who have common interests. This, too, is touted as a main feature of Delicious.

These two features – storage of personal bookmarks and the public nature of those bookmarks – are somewhat at odds with one another. The data we present below confirm that users bookmark primarily for their own benefit, not for the collective good, but may nevertheless constitute a useful public good.

## 3.1 The Data

Our analysis was performed on two sets of Delicious data, which we retrieved between the morning of Friday, June 23 and the morning of Monday June 27, 2005.

The first set ("popular") contains all the URLs which appeared on Delicious' "popular" page during that timeframe. Our dataset contains all bookmarks ever posted to each of those URLs regardless of time, so that for each URL our dataset contains its complete history within the system. A total of 212 URLs and 19,422 bookmarks comprise this dataset.

Our second dataset ("people") consists of a random sample of 229 users who posted to Delicious during the above timeframe. Our dataset contains all bookmarks ever posted by those users, regardless of time, so that for each user our dataset contains that user's complete history. A total of 68,668 bookmarks comprise this dataset.

We begin by looking at the tag use of individual users. As users bookmark new URLs, they create tags to describe them. Over time, users' lists of tags can be considered descriptive of the interests they hold as well as of their method of classifying those interests. First, we look at users' activity with respect to their tag use. Next, we examine tags themselves in greater detail.

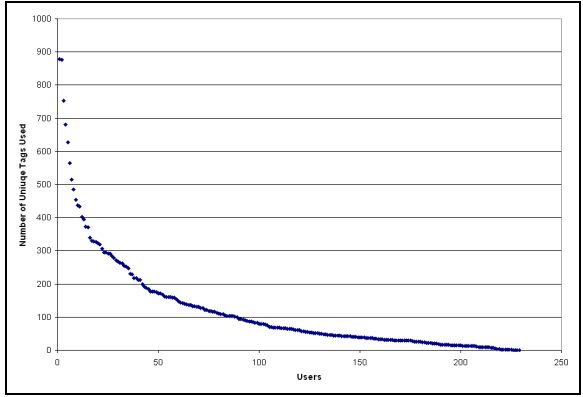

**Figure 2. The number of tags in each user's tag list, in decreasing order.**

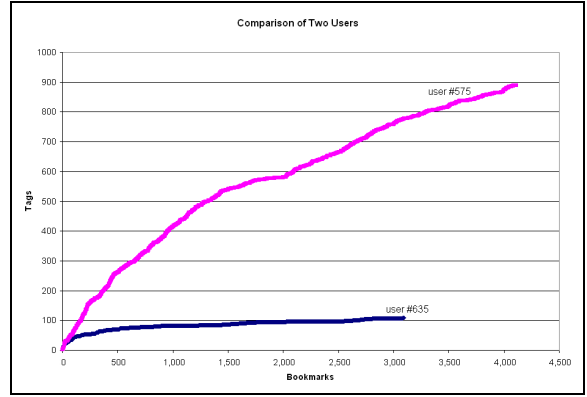

**Figure 3. Two extreme users' (#575, #635) tag growth. As they add more bookmarks, the number of tags they use increases, but at very different rates.**

## 3.2 User Activity and Tag Quantity

As might be expected, users vary greatly in the frequency and nature of their Delicious use. In our "people" dataset, there is only a weak relationship between the age of the user's account (i.e. the time since they created the account) and the number of days on which they created at least one bookmark (n=229; $R^2$=.52). That is, some users use Delicious very frequently, and others less frequently. Note that these data do not include any users who had previously used Delicious but stopped, as they were all active users at the time the dataset was collected.

More interestingly, there is not a strong relationship between the number of bookmarks a user has created and the number of tags they used in those bookmarks (n=229; $R^2$=.33). The relationship is weak at the low end of the scale, users with fewer than 30 bookmarks (n=39; $R^2$=.33), and even weaker at the upper end, users with more than 500 bookmarks (n=36; $R^2$=.14). Some users have comparatively large sets of tags, and other users have comparatively small sets (Figures 2, 3).

Users' tag lists grow over time, as they discover new interests and add new tags to categorize and describe them. Tags may exhibit very different growth rates, however, reflecting how users' interests develop and change over time. Figures 4a and 4b show how use of each tag increases as each user adds more bookmarks over time. For each user, two of those tags' usages grow steadily, reflecting continual interests tagged in a consistent way. One tag grows rapidly, reflecting a newfound interest or a change in tagging practice. It is possible that the newly growing tag represents a new interest or category to the user. Another possibility is that the user has chosen to draw a new distinction among their bookmarks, which can prove problematic for the user.

Because sensemaking is a retrospective process, information must be observed before one can establish its meaning (Weick et al. forthcoming). Therefore, a distinction may go unnoticed for a long time until it is finally created by the individual, who then continues to find that distinction important in making sense of future information. Since finding previously encountered information is extremely important (Dumais et al. 2003), this is deeply problematic for past information. For example, user # 575 (Figure 4a) did not use "tag 3" until approximately the 2500[th] bookmark. If 'tag 3" indeed constitutes a new distinction among the kinds of items this user bookmarks, though Delicious does allow users to alter previous bookmarks, it would be arduous to reconsider each of the earlier 2500 bookmarks to decide whether to add "tag 3" to them. Further, if in the future this user needs to filter his bookmarks by "tag 3", then no bookmark before the 2500th will be retrieved, compromising the practical usefulness of the tag.

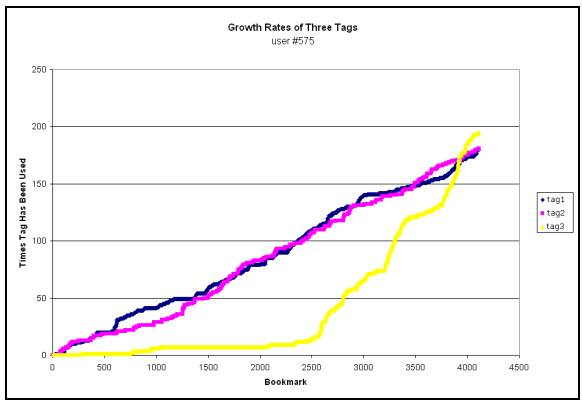 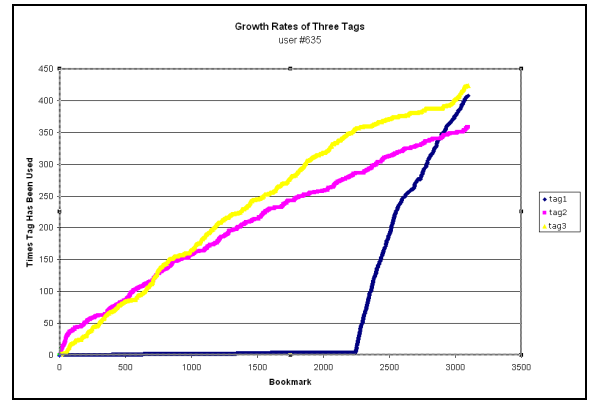

**Figure 4a,b. Growth rate of three selected tags for two users (#575, #635).**

Figures 4a and 4b show that users' tag collections, like their interests, are continually growing and evolving. Next, we look at what functions tags play in bookmarks.

### 3.3 Kinds of Tags

Tagging, as discussed above, is an act of organizing through labeling, a way of making sense of many discrete, varied items according to their meaning. By looking at those tags, we can examine what kinds of distinctions are important to taggers.

There is some discussion among the Delicious tagging community concerning whether a tag is properly considered to be descriptive of the thing itself, or descriptive of a category into which the thing falls (Coates 2005). However, we see no contradiction between these two kinds of tag. When a category is defined as circumscribing many objects with a particular property, we naturally consider each of those objects to have that property. In our estimation, the scope of the tag – whether it describes an object or a group of objects – is less interesting than the function of a tag, or what kind of information it conveys and how it is used. Here, we identify several functions tags perform for bookmarks.

1. **Identifying What (or Who) it is About**. Overwhelmingly, tags identify the topics of bookmarked items. These items include common nouns of many levels of specificity, as well as many proper nouns, in the case of content discussing people or organizations.

2. **Identifying What it Is.** Tags can identify what *kind* of thing a bookmarked item is, in addition to what it is about. For example, `article`, `blog` and `book`.

3. **Identifying Who Owns It**. Some bookmarks are tagged according to who owns or created the bookmarked content. Given the apparent popularity of weblogs among Delicious users, identifying content ownership can be particularly important.

4. **Refining Categories**. Some tags do not seem to stand alone and, rather than establish categories themselves, refine or qualify existing categories. Numbers, especially round numbers (e.g. `25`, `100`), can perform this function.

5. **Identifying Qualities or Characteristics**. Adjectives such as `scary`, `funny`, `stupid`, `inspirational` tag bookmarks according to the tagger's opinion of the content.

6. **Self Reference**. Tags beginning with "my," like `mystuff` and `mycomments` identify content in terms of its relation to the tagger.

7. **Task Organizing**. When collecting information related to performing a task, that information might be tagged according to that task, in order to group that information together. Examples include `toread`, `jobsearch`. Grouping task-related information can be an important part of organizing while performing a task (Jones et al. 2005).

The tension between tags that may be useful to the Delicious community at large and those useful only to oneself is evident here. The first three are not necessarily explicitly personal. Though identifying what some item is or is about presents some of the problems discussed earlier, like basic level differences, what unifies the first four functions is that the information is extrinsic to the tagger, so one can expect significant overlap among individuals. In contrast, the unifying characteristic of the final three functions is that the information they provide is relative to or only relevant to the tagger.

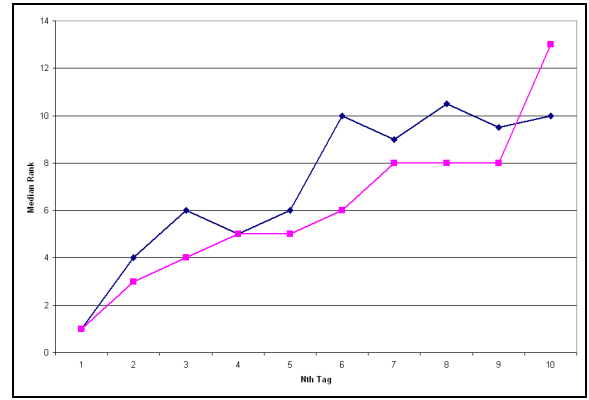

**Figure 5. As tags' order in a bookmark (horizontal) increases, its rank in the list of tags (vertical) decreases. This pattern is shown here for two URLs (#1209, #1310).**

As others have observed (Biddulph 2004), some tags are used by many people, while other tags are used by fewer people. For the reasons described above, those tags that are generally meaningful will likely be used by many taggers, while tags with personal or specialized meaning will likely be used by fewer users.

Users have a strong bias toward using general tags first. In each bookmark, the first tag used has the highest median rank (i.e. greatest frequency), and successive tags generally have a decreasing median rank (Figure 5). Earlier in the discussion of basic levels, one study showed that basic levels were those that were most quickly identified and most generally agreed upon. We suggest, therefore, that the earlier tags in a bookmark represent basic levels, because they are not only widespread in agreement, but are also the first terms that users thought of when tagging the URLs in question. A system seeking to make use of this data in order to establish the most broad categories might therefore not only look at the tags that are overall most popular, but also at those that are used earliest within bookmarks.

## 4. BOOKMARKS

We turn our attention to URLs, the bookmarks that reference them, and the tags that describe them. Here we look at how URLs are bookmarked over time, and at how the sets of tags in a URL's bookmarks constitute a stable way of describing a URL's content.

### 4.1 Trends in Bookmarking

It has been observed elsewhere (Biddulph 2004) that URLs often receive most of their bookmarks very quickly, the rate of new bookmarks decreasing over time. While true, this tells only part of the story. While many URLs (e.g. Figure 6a) do indeed reach their peak of popularity as soon as they reach Delicious, many other URLs (e.g. Figure 6b) have relatively few bookmarks for a long time until they are "rediscovered" and then experience a rapid jump in popularity. Of the 212 popular URLs in our dataset, 142 (67%) reached their peak popularity in their first 10 days in Delicious, 37 of which (17%) on their first day. However, at the other end of the spectrum, another 37 (17%) were in the

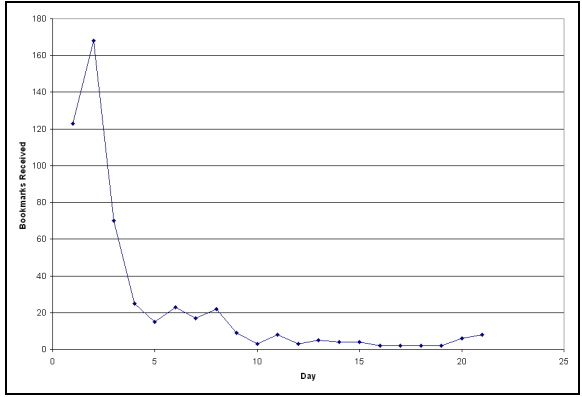
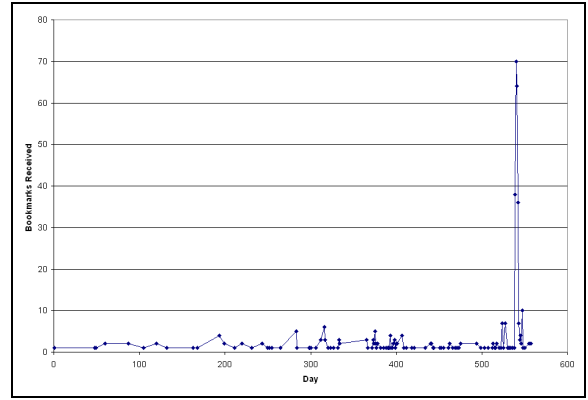

**Figure 6a,b. The addition of bookmarks to two URLs (#1310, #1209) over time. URL #1310 (left) peaks immediately, whereas #1209 (right) is long obscure before peaking.**

Delicious system for at least six months before reaching their peak of popularity. The URL in our sample that took the longest amount of time to peak in popularity did not do so until it had been in the system for over 33 months.

A burst in popularity may be self-sustaining, as popular URLs are displayed on the "popular" page, which users can visit to learn what others are currently talking about. However, the initial cause of a popularity burst is likely exogenous to Delicious; given that Delicious is a bookmarking service, a mention on a widely read weblog or website is a plausible primary cause. Kumar et al. (2003) demonstrate "burstiness" among links in weblogs, and literature on opinion and fad formation demonstrate how "well-connected" individuals and "fashion leaders" can spread information and influence others (Wu & Huberman 2005; Bikhchandani, Hirshleifer & Welch 1998).

## 4.2 Stable Patterns in Tag Proportions

As a URL receives more and more bookmarks, the set of tags used in those bookmarks, as well as the frequency of each tag's use within that set, represents the combined description of that URL by many users.

One might expect that individuals' varying tag collections and personal preferences, compounded by an ever-increasing number of users, would yield a chaotic pattern of tags. However, it turns out that the combined tags of many users' bookmarks give rise to a stable pattern in which the proportions of each tag are nearly fixed. Empirically, we found that, usually after the first 100 or so bookmarks, each tag's frequency is a nearly fixed proportion of the total frequency of all tags used. Figures 7a and 7b show this pattern. Each line represents a tag; as more bookmarks are added (horizontal axis), the proportion of the tags represented by that tag (vertical axis) flattens out. A web tool that visualizes Delicious data, called Cloudalicious also shows this pattern.

This stable pattern can be explained by resorting to the dynamics of a stochastic urn model originally proposed by Eggenberger and Polya to model disease contagion (Eggenberger & Polya 1923). In its simplest form, this probabilistic model consists of an urn initially containing two balls, one red and one black. At each time step, a ball is randomly selected and replaced in the urn along with an additional ball of the same color. Thus, after N steps, the urn contains N+2 balls. The remarkable property of such a model is that, in spite of its random nature, after a number of draws a pattern emerges such that the fraction of balls of a given color becomes stable over time. Furthermore, that fraction converges to a random limit. This implies that if the process is run forever the fraction converges to a limit, but the next time one starts the process over and run it again the stable fraction will converge to a different number.

This behavior is shown in Figures 7a and 7b, which are indistinguishable from what one would obtain from running a computer simulation of the urn model, where the colored balls would correspond to the tags observed.

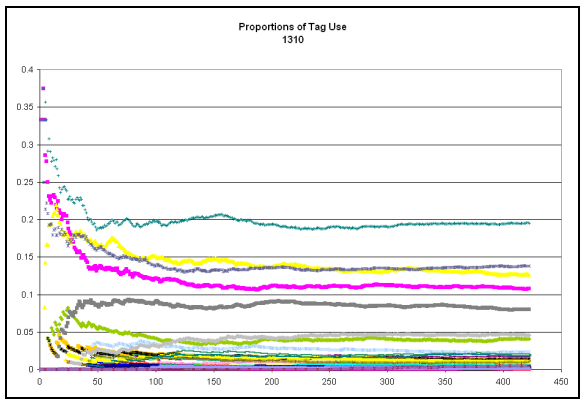
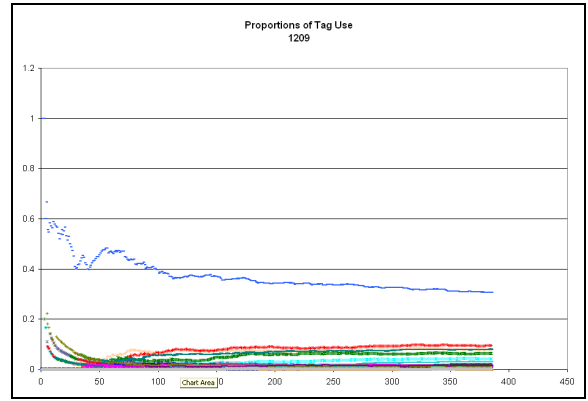

**Figure 7a,b. The stabilization of tags' relative proportions for two popular URLs (#1310, #1209). The vertical axis denotes fractions and the horizontal axis time in units of bookmarks added.**

This stability has important implications for the collective usefulness of individual tagging behavior. After a relatively small number of bookmarks, a nascent consensus seems to form, one that is not affected by the addition of further tags. Users may continually add bookmarks, but the stability of the overall system is not significantly changed. The commonly used tags, which are more general, have higher proportions, and the varied, personally-oriented tags that users may use can coexist with them.

Moreover, because this stability emerges after fewer than 100 bookmarks, URLs need not become very popular for the tag data to be useful. For example, Figures 6b and 7b show URL #1209, which reaches its stable pattern at least 100 days before its spike in popularity.

Two reasons why this stabilization might occur are imitation and shared knowledge. In the probabilistic model, replacement of a ball with another ball of the same color can be seen as a kind of imitation. Likewise, Delicious users may imitate the tag selection of other users. The Delicious interface through which users add bookmarks shows users the tags most commonly used by others who bookmarked that URL already; users can easily select those tags for use in their own bookmarks, thus imitating the choices of previous users. This can be helpful, especially if a user does not know how to categorize a particular URL. A user may use the suggested popular tags as a way of looking to others to see what the "right" thing to do is. The principle of "social proof" suggests that actions are viewed as correct to the extent that one sees others doing them (Cialdini 2001). In this case, choosing tags others have already used may seem like a "safe" choice, or one that does not require time or effort.

Imitation, however, does not explain everything. The interface of Delicious shows only a few of the most commonly used tags, but the stable pattern persists even for less common tags, which are not shown. Shared knowledge among taggers may also account for their making the same choices.

Recall from above the social aspect of sensemaking. We may expect that users of Delicious, or any other tagging system, share some background, linguistic, cultural, educational, and so on. In the case of Delicious, many users appear to have strong technical backgrounds, as many of the bookmarked URLs are technology-related. Accordingly, some documents may occupy roughly the same status in many of those users' lives; since they may make use of web documents in the same way, users may categorize them the same way, as well.

Part of the reason these stable patterns emerge is that the ideas and characteristics that are represented in tags are stable. As the ideas themselves change, these stable states may likewise change. For example, in Speroni (2005), one URL's Cloudalicious chart shows surging growth of the tag "ajax." This example illustrates a change in how some users categorized the technologies they use in their work. Specifically, the term "ajax" came to represent a set of technologies that previously had no name, but as this concept came into being, it represented an innovation that changed the way some people thought and spoke about the technologies they used, as well as the URLs they tagged about those technologies.

## 5. CONCLUSION
We have observed that collaborative tagging users exhibit a great variety in their sets of tags; some users have many tags, and others have few. Tags themselves vary in frequency of use, as well as in what they describe. Nevertheless, because stable patterns emerge in tag proportions, minority opinions can coexist alongside extremely popular ones without disrupting the nearly stable consensus choices made by many users.

The prevalence of tagging with a very large number of tags and according to information intrinsic to the tagger demonstrates that a significant amount of tagging, if not all, is done for personal use rather than public benefit. Nevertheless, even information tagged for personal use can benefit other users. For example, if many users find something `funny`, there is a reasonable likelihood someone else would also find it to be so, and may want to explore it. Likewise, one might want to read something that many other people have decided they want `toread` as well. In this way, Delicious functions as a recommendation system, even without explicitly providing recommendations. However, information tagged by others is only useful to the extent that the users in question make sense of the content in the same way, so as to overlap in their classification choices.

The stable, consensus choices that emerge may be used on a large scale to describe and organize how web documents interact with one another. Currently this is being performed, problematically, on small scales by experts and, equally problematically, on large scales by machines (Shirky 2005; Doan et al. 2002). The stability we have shown here demonstrates that tagged bookmarks may be valuable in aggregate as well as individually, in performing this larger function across the web.

Given the current proliferation of sites that support collaborative tagging, we expect that these sites will continue to provide a fertile ground for studying computer mediated collaborative systems in addition to providing users with new ways to share and organize content.

## 6. REFERENCES

Biddulph, M. (2004). Introducing Del.icio.us. *XML.com*. *http://www.xml.com/pub/a/2004/11/10/delicious.html*.

Bikhchandani, S., Hirshleifer, D. & Welch, I. (1998). Learning from the Behavior of Others: Conformity, Fads and Informational Cascades. *Journal of Economic Perspectives*. 12(3). 151-170.

Cialdini, R. (2001). *Influence: Science and Practice.* 4[th] ed. Allyn and Bacon.

CiteULike. *http://www.citeulike.org/*

Cloudalicious. *http://cloudalicio.us/*

Coates, T. (2005). Two Cultures of Fauxonomies Collide. *http://plasticbag.org*. June 4, 2005.

Connotea. *http://www.connotea.org/*

Del.icio.us. *http://del.icio.us/*

Doan, A., Madhavan, J., Domingos, P. & Halevy, A. (2002). Learning to Map between Ontologies on the Semantic Web. *Proceedings of WWW2002*.

Dumais, S., Cuttrell, E., Cadiz, J., Jancke, G., Sarin, R., & Robbins, D. (2003). Stuff I've Seen: A System for Personal Information Retrieval and Re-Use. *Proceedings of the ACM Conference on Information Retrieval*.

Eggenberger, F. and Polya, G. (1923). Uber die Statistik verketter vorgage. *Zeit. Angew. Math. Mech.* 1, 279-289.

Flickr. *http://www.flickr.com/*

Jones, W., Phuwanartnurak, A., Gill, R. & Bruce, H. (2005). Don't Take My Folders Away! Organizing Personal Information to Get Things Done. *Proceedings of the ACM Conference on Human Factors in Computing Systems (CHI)*.

Kumar, R., Novak, J., Raghavan, P. & Tomkins, A. (2003). On the Bursty Evolution of Blogspace. *Proceedings of WWW2003*.

Labov, W. (1973). The Boundaries of Words and their Meanings. In C.J. Bailey and R. Shy, eds., *New Ways of Analyzing Variation in English*.

Mathes, A. (2004). Folksonomies – Cooperative Classification and Communication Through Shared Metadata. *http://www.adammathes.com/academic/computer-mediated-communication/folksonomies.html*

Pustejovsky, J. (1995). *The Generative Lexicon*. MIT Press.

Rowley, J. (1995). *Organizing Knowledge*. 2nd Ed. Brookfield, VT: Gower.

Shirky, C. (2005). Ontology is Overrated: Categories, Links and Tags. *http://www.shirky.com/writings/ontology_overrated.html*

Speroni, P. (2005). Tagclouds and Cultural Changes. *http://blog.pietrosperoni.it*. May 28, 2005.

Tanaka, J., & Taylor, M. (1991). Object Categories and Expertise: Is the Basic Level in the Eye of the Beholder? *Cognitive Psychology* 23(3). 457-482.

Technorati. *http://www.technorati.com/*

Weick, K., Sutcliffe, K. & Obstfeld, D. (forthcoming). Organizing and the Process of Sensemaking. *Organizational Science*.

Wu, F. & Huberman, B. (2005). Social Structure and Opinion Formation. *http://www.hpl.hp.com/research/idl/papers/opinions/opinions.pdf*

Yahoo MyWeb 2.0. *http://myweb2.search.yahoo.com/myweb*